\documentclass[12pt,preprint,osajnl,showpacs]{revtex4}
\usepackage{psfig}
\usepackage{graphicx}

\newcommand{\bgar}{\begin{eqnarray}}
\newcommand{\enar}{\end{eqnarray}}
\newcommand{\eq}[1]{(\ref{eq:#1})}

\newcommand{\eqname}[1]{\label{eq:#1}}

\newcommand{\al}[1]{^{(#1)}}

\begin{document}
 \title{\bf Cavity-enhanced single frequency synthesis via DFG of
 mode-locked pulse trains}

\author{Gabriele Ferrari}
 \address{INFM, European Laboratory for Non-linear Spectroscopy,
 via N. Carrara 1, I-50019 Sesto Fiorentino, Italy.}

 \email{ferrari@lens.unifi.it}

 \author{Iacopo Carusotto}
 \address{Laboratoire Kastler Brossel, \'Ecole Normale Sup\'erieure, 24 rue
 Lhomond, F-75231 Paris Cedex 05, France}

\address{BEC-INFM Research and Development Center, I-38050 Povo, Trento,
Italy}

\email{carusott@science.unitn.it}

\begin{abstract} We show how to synthesize a CW, single-frequency optical
field from the frequency-dispersed, pulsed field of a mode-locked laser. This
process, which relies on difference frequency generation in an optical cavity,
is efficient and can be considered as an optical rectification. Quantitative
estimates for the output power and amplitude noise properties of a realistic
system are given. Possible applications to optical frequency synthesis and
optical metrology are envisaged.

\end{abstract}


\maketitle

\section{Introduction}

The introduction of Kerr-lens mode-locked  lasers (KLML) as referenced optical
frequency synthesizers \cite{Diddams2000,Udem2002} represents a milestone in
optical frequency synthesis. Before the advent of KLML, optical frequency
synthesis and measurements were generally performed through harmonic
generation in frequency chains \cite{Jennings1983,Schnatz1996} relying on
involved apparatuses. In addition, only  narrow portions of the optical
spectrum were accessible. Alternative techniques of frequency interval
bisection \cite{Telle1990,Udem1997} and optical comb generators
\cite{Kourogi1993,Kourogi1996} simplified to some extent the task of
synthesizing optical frequencies, but still could not provide a single
instrument to cover the whole optical spectrum from the visible to the near
infra-red.

By providing a wide ruler of well spaced optical frequencies, KLML now
represent the best solution to measure and synthesize optical frequencies
ranging from the near IR to most of the visible. These lasers have a spectrum
composed by a comb of frequencies $f_{m}=m\,f_{\rm r}+f_{\rm CEO}$, where
$f_{\rm r}$ is the repetition rate of the pulsed laser, $f_{\rm CEO}$ is the
carrier envelope offset frequency, and $m$ is an integer \cite{Cundiff2001}. On
one hand, optical frequency measurements are performed by measuring the beat
note between the optical frequency $f$ to be measured and the closest tooth of
the KLML comb, and then determining the integer $m$ from a rough wavelength
measure. On the other hand, an optical frequency can be synthesized by locking
the frequency of a continuous wave optical field to a tooth of the KLML comb.

From the practical point of view, such optical measurements depend critically
on two issues: the determination and stabilization of the comb parameters
$f_{\rm r}$ and $f_{\rm CEO}$, and the amount of power available at the
frequency component $f_{m}$ closest to the one of the laser field to be
measured \cite{Hollberg2001}. While $f_{\rm r}$ is directly available from the
amplitude modulation of the KLML, the determination and stabilization of
$f_{\rm CEO}$, typically known as {\em self-referencing}, involves
delicate manipulation of the ultra-broadband light, e.g. by means of nonlinear
optical processes such as frequency doubling of
the red tail of the spectrum and its beating with the blue tail
\cite{Cundiff2001} or even more complex schemes, as discussed
in~\cite{Telle1999}.

This, in turn, requires the use of octave-spanning sources
such as Ti:Sa femto-second lasers spectrally broadened through micro-structured
fibers \cite{Ranka2000}, or ultra-broadband femto-second lasers
\cite{Ramond2002}. From the point of view of the power available on the
frequency component $f_{m}$ of interest, self-referenced comb generators were
demonstrated to have a suitable spectrum to measure frequencies in the
wavelength range from 500\,nm up to 2\,$\mu$m \cite{Cundiff2001,Thomann2003}.
The region of optimal sensitivity for the measurement depends on the frequency
comb spectral envelope and therefore on the nature of the medium used to
generate the ultra-broadband spectrum \cite{Hollberg2001}.

\section{Cavity-assisted rectification: general idea}

We propose a different approach to directly synthesize optical fields from
mode-locked lasers. The basic idea consists in the conversion of the comb of
frequency components present in a mode-locked laser into a single frequency
component $\bar{f}$. The idea can be implemented to devise an optical rectifier
that efficiently converts a pulsed light source into a monochromatic continuous
wave optical field. Two regions of the spectrum of a mode-locked laser are sent
through a non-linear crystal so as to generate radiation via difference
frequency generation (DFG). The nonlinear polarization in the crystal is
composed by a comb of frequency which are integer multiples of the repetition
rate $f_{\rm r}$. An optical cavity surrounds the non-linear crystal and is
resonant with the polarization component at frequency $\bar{f}$, while all
other components are off-resonance with respect to all cavity modes. In these
conditions, the polarization mode at $\bar{f}$ solely contributes to the field
circulating in the cavity, hence a nearly monochromatic radiation is generated
with a power proportional to the finesse of the cavity. This results in a power
on the component at $\bar{f}$ orders of magnitude higher than the one that
would be obtained by means of a simple frequency filtering of the radiation
generated by a single-pass DFG. Moreover, the suppression of all non resonant
components by the action of the cavity results in a significant reduction of
the amplitude noise spectrum.

\section{Theoretical model}

\begin{figure}[htbp]
\begin{center}   \includegraphics[clip,width=6cm]{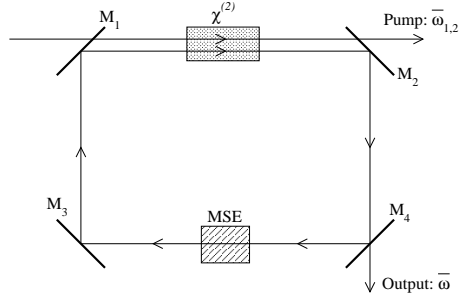}
\caption{Experimental setup of the optical rectifier for mode-locked lasers. A
non-linear medium, chosen for difference frequency generation (DFG) between two
spectral regions of a mode-locked laser, is placed into an optical cavity
resonant with one frequency component ${\bar \omega}$ generated by DFG. The
$M_{1,2,3}$ mirrors are highly reflective at the frequencies generated by DFG
and highly transmitting for all the other frequencies. The output mirror $M_4$
is chosen to be slightly transmitting for the circulating light. MSE: mode
selective etalons.} \label{fig:Setup}
\end{center}
\end{figure}

Consider a ring cavity of total length $L$ containing a slab of nonlinear
medium of length $\ell$, as sketched in fig.\ref{fig:Setup}.
The cavity mirrors are assumed to be perfectly transmitting for the
source laser light so as to avoid the need of matching the cavity
spectral range with the comb spacing,
but highly reflecting at the ${\bar \omega}$ of the radiation to be
generated so as to take advantage of resonance effects at ${\bar \omega}$.
Here, the reflectivity $R$ of three mirrors will be taken exactly 1,
while the fourth one will have a finite but small transmission $T \ll
1$ and will be considered as the output of the system.

The source beam consists of a train of ultrashort pulses
\begin{equation}
E(t)=\sum_n E_0(t-n\tau)\, e^{in\phi_0},
\eqname{PulseTrainT}
\end{equation}
where $E_0(t)$ describes the shape of each pulse, $\tau$ is the temporal
spacing of the pulses and $\phi_0$ is their relative phase. In Fourier space
this train of pulses corresponds to a comb of $\delta$-function peaks
\begin{equation}
 {\tilde E}(\omega)=2\pi\, {\tilde
  E}_0(\omega)\,\sum_{n=-\infty}^\infty \delta(\omega \tau+\phi_0-2\pi n),
\eqname{PulseTrainOmega}
\end{equation}
at equispaced frequencies $\omega_m=m\,\omega_{\rm r}+\omega_0$, where
$\omega_{\rm r}=2\pi/\tau=2\pi\,f_{\rm r}$ is the comb spacing and
$\omega_0=-\phi_0/\tau=2\pi\,f_{\rm CEO}$ is the carrier envelope offset. The
comb envelope ${\tilde E}_0(\omega)$ is the Fourier transform  of the single
pulse profile $E_0(t)$, which is centered at the carrier frequency
$\omega_c$ and has a width of the order of the inverse of the single
pulse duration $t_{\rm p}$.

Since we are interested in generating light at a given frequency ${\bar
f}={\bar \omega}/2\pi$, two spectral sections of the comb are selected centered
at ${\bar \omega_1}$ and ${\bar \omega_2}$, respectively, and such that their
difference ${\bar \omega_1}-{\bar \omega_2}$ is close to ${\bar
  \omega}$.
In the following, we shall see that the width $\Delta\omega$ of the
two sections is limited by the acceptance bandwidth $\Delta\omega_{\rm
  acc}$ of the nonlinear crystal as well as by the different group
velocity of the two sub-pulses at ${\bar \omega}_{1,2}$ during the
propagation towards the crystal, i.e. the dephasing of the different
teeth within each of the two sections.
Concerning the latter point, dispersion can be compensated with
standard techniques such as chirped mirrors and/or negatively
dispersing optical elements, while for the dispersion in the nonlinear
crystal the acceptance bandwidth is directly obtained from data
available in literature. We will discuss in the following a possible
experimental realization.

We start by assuming a constant nonlinear susceptibility of the crystal equal
to $\chi\al{2}$, and we neglect source depletion and reflections at the crystal
interfaces.
As the relative phase between the different teeth within each of the
two sections around
${\bar \omega}_{1,2}$ is assumed to be negligible, the nonlinear
polarization inside
the crystal can be written as:
\begin{equation}
P_2(z,t)=\chi\al{2}\,\sum_{j,m}{\tilde E}^*(\omega_{m})\,{\tilde
  E}(\omega_{j})\,e^{i(k_{j}-k_{m})z}
\,e^{-i(\omega_{j}-\omega_{m})t},
\eqname{P_2Omega}
\end{equation}
where
$n(\omega_j)=\sqrt{\epsilon(\omega_j)}$ is the linear refractive index of the
nonlinear medium at $\omega_j$ and $k_j=n(\omega_j)\,\omega_j/c$ is the
corresponding wave vector; $n$ is assumed to be real at all frequencies of
interest; the indices $j$ and $m$ run over the frequency components of the
comb. The Fourier transform of \eq{P_2Omega} is a comb of $\delta$-function
peaks spaced of $\omega_{\rm r}$ with a vanishing offset. The amplitude of the
component at the frequency $\omega_p=p\,\omega_{\rm r}$ is given by
the sum over all pairs $(j,m)$ such that $j-m=p$.

The steady-state amplitude of the cavity field is obtained by solving the wave equation with the
source term corresponding to the nonlinear polarization \eq{P_2Omega}. As the calculation is
somewhat lengthy, the details are given in the Appendix~\ref{sec:appendix}, and we here discuss
the final result only. The frequency spectrum of the cavity field has the same discrete structure
of the nonlinear polarization. On the internal side of the output mirror, the amplitude of the
$\omega=\omega_p$ component has the explicit expression:
\begin{eqnarray}
  \label{eq:ElectrField0}
{\tilde E}_{\rm int}(\omega_p)= &&\frac{1}{(1-T)^{1/2} -e^{-i \omega_p L_{\rm eff}/c}}
\,e^{ik_{NL}z_1} \,e^{-i\omega_p z_1/c}
\\
&& \frac{4\pi\omega_p^2\,n(\omega_p)^{1/2}}{c^2} \nonumber \\
&&\sum_{j-m=p}  \frac{e^{i [k_{\rm NL}^{(j,m)}-k_p]\ell}-1}
 {k_p^2-k_{\rm NL}^{(j,m)\;2}}\,\chi\al{2}\,E^*(\omega_{m})\,E(\omega_{j}), \nonumber
\end{eqnarray}
where the optical length of the cavity is defined as usual as $L_{\rm
eff}=n(\omega_p)\,\ell+(L-\ell)$ and $z_1$ gives the position of the front
interface of the crystal.

The denominator of the first fraction on the RHS accounts for resonance
effects due to the
presence of the cavity whose modes are determined by the usual round-trip
condition $\omega_q L_{\rm eff}/c = 2 \pi q$ ($q$ integer). For a weak
transmittivity $T\ll 1$ of the output mirror, and in the absence of
significant losses in the cavity, the damping rate $\Gamma$ of the cavity modes
is determined by the finite transmittivity $T$ of the output mirror
$\Gamma\approx\Gamma_{\rm rad}=Tc/L_{\rm eff}$. The fraction in the sum takes
into account the phase-matching effects associated with the refractive index
dispersion of the nonlinear crystal. Because of dispersion, the polarization at
$\omega_p$ produced by the different pairs $(j,m)$ has a different wave vector
$k^{(j,m)}_{\rm NL}=k_j-k_m$. This wave vector has to be close to the one of
the radiation to be generated $k_p=n(\omega_p)\,\omega_p/c$ for the DFG process
to be efficient.

Physical insight on the spectrum of the generated light can be found by rewriting the amplitude of
the output beam ${\tilde E}_{\rm out}(\omega_p)=\sqrt{T}\,{\tilde  E}_{\rm int}(\omega_p)$ in
terms of the field ${\tilde E}_{\rm sp}(\omega_p)$ that would be generated by the same crystal in
a single-pass geometry. The same calculations (see Appendix~\ref{sec:appendix}) that led to
\eq{ElectrField0} give the following  expression:
\begin{equation}
  \label{eq:ElectrFieldT2}
{\tilde E}_{\rm out}(\omega_p)=\frac{\sqrt{T}}{1-(1-T)^{1/2}\,e^{i\omega_p
    L_{\rm eff}/c}}\,{\tilde E}_{\rm
    sp}(\omega_p).
\end{equation}
For $T\ll 1$, one sees that the output power which is obtained under the
resonance condition is a factor $4/T$
larger than the one that would
be obtained in a single-pass geometry. If the non-radiative cavity losses
$\Gamma_{\rm loss}$ are not negligible, the output power is a factor
$1/\big(1+\Gamma_{\rm loss}/\Gamma_{\rm rad}\big)^2$ weaker.

\section{Illustrative examples}

As we are interested in obtaining a single-frequency output beam, a single
tooth out of the comb of frequencies actually present in the nonlinear
polarization spectrum has to be singled out by the action of the cavity. This
has therefore to be designed in such a way that only the component at the
desired frequency ${\bar \omega}$ is on resonance with a cavity mode, while all
the others, being off-resonance, are suppressed. To this purpose, several
strategies can be adopted. The most trivial one consists in choosing  the
cavity free spectral range and the repetition rate of the source laser to be
non-commensurate to avoid multiple exact resonances \cite{HighFinesse}. Another
option is, in analogy with standard laser techniques, to insert in the cavity
some frequency discriminators, like Fabry-Perot etalons, which introduce
additional losses to suppress unwanted resonances keeping the resonant
enhancement only for the mode of interest.

\begin{figure}[htbp]
\begin{center}   \includegraphics[clip,width=9cm]{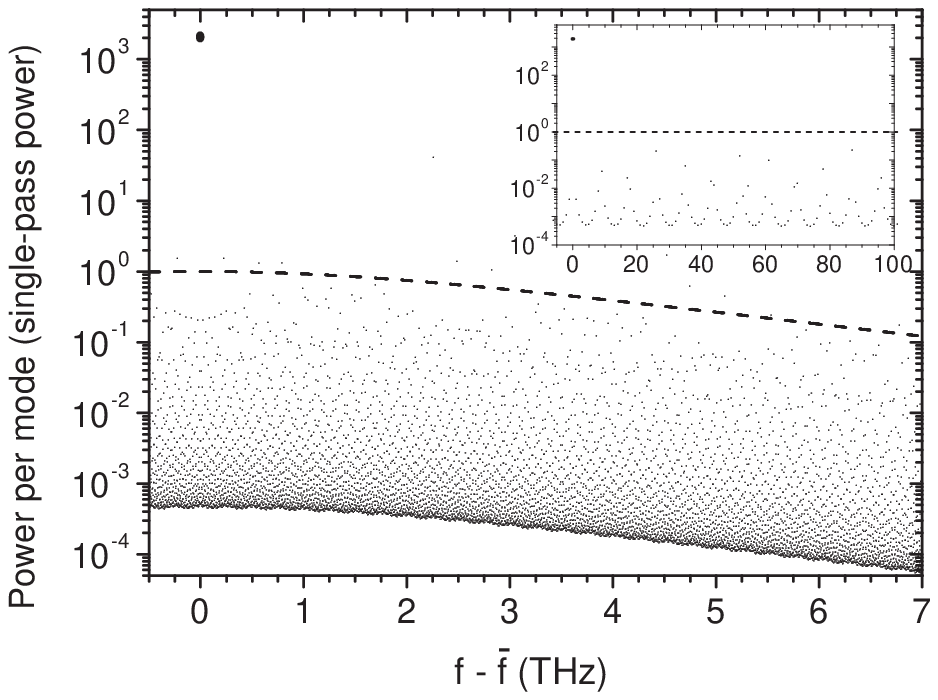}
\caption{Output power-spectrum of the rectifier discussed in the
  text. Each point corresponds to the frequency component of the
  polarization at $f_p=\omega_p/2\pi$. The power is normalized to
the single-pass conversion efficiency of the nonlinear crystal at the center
frequency ${\bar f}$; the dashed line corresponds to the emission that
one would have in a single-pass DFG.
The inset represents a magnification of the same spectrum around the
resonant mode (bold circle), with the frequency expressed in GHz.
The polarization is here considered to have a $1/\cosh$ profile
  centered at ${\bar f}$ with FWHM of 6.4\,THz (dashed line).
The graph shows the increase of the power available on the resonant
frequency component and the corresponding reduction of the power on all the
  non-resonant components.}
  \label{fig:Spectrum}
\end{center}
\end{figure}

As an illustrative example, we consider a resonator with a free spectral range
(FSR) of 1,13\,GHz pumped by a femtosecond laser with a repetition rate $f_{\rm
r}$ of 1\,GHz and containing two uncoated glass etalons 410\,$\mu$m and 2.2\,mm
thick. The two etalons have unity transmission at $\bar{f}$, and the finesse of
the cavity is 1570 and limited by the transmissivity of the output mirror
$T_{\rm out}$=0.2\%. The polarization of the non-linear crystal is a $1/\cosh$
profile with FWHM of 6.4\,THz and centered on $\bar{f}$. Figure
\ref{fig:Spectrum} represents the output spectrum for these specific parameter
values with the optical power normalized to field at $\bar{f}$ generated in
single-pass DFG. The inset shows that the resonator not only enhances the
component at $\bar{f}$ but also suppresses the power on most of the other,
non-resonant, components. With the parameters considered here, the power at
$\bar{f}$ is increased by a factor 2000 with respect to the single-pass
conversion, and the sum of the power on all non-resonant modes amounts to
13\,\% of that at $\bar{f}$. This means that the output is indeed a
continuous wave field at frequency $\bar{f}$ superposed to a small,
amplitude modulation at frequency $f_{\rm r}$ and its higher harmonics
with a bandwidth of the order of the frequency spread of the DFG
polarization, of the order of $\Delta\omega/2\pi$.

Giving a realistic estimate of the power generated by the optical rectifier
requires stringent assumptions on the experimental parameters
\cite{ExperimentalAssumptions}. We examine, for instance, the case in which the
optical rectifier synthesizes radiation at 2.8\,$\mu$m. We consider the
spectral windows at 680 and 900\,nm form a mode-locked laser, each having a
spectral width of 3\,THz equal to the acceptance bandwidth of the non-linear
crystal. Supposing that the model-locked laser has 1\,GHz repetition rate and
delivers both at 680 and 900\,nm 30\,pJ per pulse \cite{Ramond2002}, that the
non-linear medium is a 1\,mm crystal of periodically poled lithium niobate
(PPLN), neglecting losses \cite{Myers1996}, then the single-pass DFG energy can
be estimated to be of 0.2\,pJ per pulse. With the given  repetition rate
$f_{\rm r}$=1\,GHz and spectral width for the envelope of the DFG radiation,
the peak single-pass conversion to 2.8\,$\mu$m is about 35\,nW per frequency
component. The use of the optical rectifier increases the output power at
$\bar{f}$ to 70\,$\mu$W. This value is comparable \cite{Borri2003} or larger
\cite{Fuji2004} than the power produced by alternative systems with analogous
spectral characteristics.

\section{Discussion}

Let's now summarize the main advantages offered by the proposed optical
rectifier as a frequency synthesizer of continuous wave optical fields. It is
well known that the frequency components generated by DFG between separate
spectral regions of one mode-locked laser depend only on the repetition rate
$f_{\rm r}$ and not on the carrier envelope offset frequency $f_{\rm CEO}$.
This makes its stabilization unnecessary and therefore reduces the complexity
of the system.

Another common feature of DFG processes with mode-locked lasers is that the
amplitude at the frequency ${\bar f}$ is not determined by a single $(l,m)$
pair, but from the constructive interference of all the pairs $l-m=p$ that lie
within the acceptance window of the nonlinear crystal $\Delta \omega_{\rm
acc}$. If we denote by $(l^*,m^*)$ the pair of components which gives exact
phase matching $k_{NL}^{(l^*,m^*)}=n({\bar \omega})\,{\bar \omega}\,/c$ and we
linearize the dispersion of the refractive index around its values respectively
at $\omega_{l^*}$ and $\omega_{m^*}$, the number $N_c$ of pairs which are
actually phase matched can be estimated to be of the order of:
 \begin{equation}
    \label{eq:N_c}
   N_c\approx\frac{\pi}{\ell\,\omega_{\rm r}}\;\Big|\frac{1}{v_g({\bar
       \omega}_{1})}-
 \frac{1}{v_g({\bar \omega}_{2})}\Big|^{-1}=\frac{\Delta \omega_{\rm
       acc}}{\omega_{\rm r}}.
  \end{equation}
The power of the generated beam is therefore a factor $N_c^2$ larger than the
one that would be generated by a single
 $(l,m)$ pair.

In our cavity-enhanced set-up, the power of the generated light is further
enhanced by a factor proportional to the finesse. Moreover, suppressing the
power on all frequency components available from the polarization  but
$\bar{f}$, the optical rectifier allows for a strong reduction on the amplitude
modulation of the generated field. This results into a reduction of the noise
spectrum of the beatnote signal between the synthesized frequency and that to
be measured \cite{Hollberg2001}. This should lead to high contrast measurements
even with relatively weak fields.

In conventional lasers the coherence properties of the generated light are
determined by the spontaneous symmetry breaking induced by coupling between the
cavity and the gain medium \cite{LaserRef}. However, it is worth noting that in
our optical rectifier the coherence properties of the output are determined by
the ones of the mode-locked laser and, in particular, through its repetition
rate. The cavity plays only the role of selector among the pairs of frequencies
allowed by difference frequency generation, enhancing one single frequency
while suppressing the others.

From the practical point of view, the spectrum of the pumping mode-locked laser
has to be taken into account in the choice of the cavity parameters. Neglecting
all cavity losses but the output mirror, equation (\ref{eq:ElectrFieldT2})
shows that output power increases as $1/T$. This holds true in the
most usual case when the linewidth of each frequency component of the
mode-locked laser is narrower than the
linewidth of the cavity mode. By increasing the finesse of the cavity beyond
this condition one can easily prove that the circulating power now increases
proportionally to 1/T, while the output power tends to a constant.

\section{Conclusions}

In conclusion, we presented a novel scheme for synthesizing continuous wave and
single-frequency optical beams from pulsed mode-locked lasers. We believe that
this scheme for optical rectification holds promises for applications to
optical frequency metrology and direct synthesis of CW optical fields from
radiofrequencies. The continuous wave optical synthesizer would be composed by
a mode-locked laser spanning the spectral regions to pump the DFG, with the
repetition rate stabilized to a radiofrequency reference, followed by the
optical rectifier tuned to the frequency to be generated.

\section*{Acknowledgments}

We thank M.\,Artoni, R.E.\,Drullinger, G.\,Giusfredi, D.\,Mazzotti, and
G.M.\,Tino for stimulating discussions. Laboratoire Kastler Brossel is a
unit\'e de Recherche de l'Ecole Normale Sup\'erieure et de l'Universit\'e
Pierre et Marie Curie, associ\'ee au CNRS.

\appendix

\section{Solution of the wave equation}
\label{sec:appendix}
The purpose of this appendix is to give some details about the calculations
that lead to the expression \eq{ElectrField0} for the in-cavity field
generated by the nonlinear polarization of the medium.

As Maxwell's equations are linear, we can calculate the emitted field
separately for each pair $(j,m)$ such that $\omega_j-\omega_m=\omega_p$ which
then contributes to the polarization at the frequency $\omega_p$ of
interest. The final result \eq{ElectrField0} will be the sum of all these
contibutions. From now on all quantities are to be considered as Fourier
components at the frequency $\omega_p$ of the corresponding quantity.
The spatial coordinate $z$ goes around the cavity; let $z=0$ correspond to the
output mirror, and the nonlinear medium extend from $z=z_1$ to $z=z_1+\ell$.

The general solution of Maxwell's equation in the free-space outside the
nonlinear medium has the usual plane-wave form:
\begin{equation}
E_{\rm fs}(z)={\bar E}_{\rm fs}\,e^{i \omega_p z/c}.
\end{equation}
There is no need for taking into account the counterpropagating wave thanks to the assumption of
negligible reflections at the nonlinear medium interfaces. The general solution \eq{wave_fs} has
to be applied in the two regions outside the crystal $z\in[0,z_1]$ and $z\in[z_1+\ell,L]$, where
the field can be respectively written as
\begin{equation}
E_{\rm fs}\al{1,2}(z)={\bar E}_{\rm fs}\al{1,2}\,e^{i \omega_p z/c} \eqname{wave_fs}
\end{equation}
${\bar  E}_{\rm fs}\al{1,2}$ being parameters to be determined later.

Using the explicit expression for nonlinear polarization of the nonlinear
medium:
\begin{equation}
P_2(z)=P_{NL}\,e^{ik_{NL}z},
\eqname{P_2Omega_1}
\end{equation}
the inhomogeneous wave equation inside the nonlinear medium has the form:
\begin{equation}
\left(\frac{\partial^2}{\partial z^2}+
\epsilon(\omega_p)\,\frac{\omega_p^2}{c^2}\right)E(z)=
-\frac{4\pi\omega_p^2}{c^2}\,P_{NL}\,e^{ik_{NL}z}.
\end{equation}
where $k_{NL}=k_j-k_m$ gives the spatial dependence of the nonlinear
  polarization, $P_{NL}=\chi\al{2}\,{\tilde E}^*(\omega_{m})\,{\tilde
  E}(\omega_{j})$ its amplitude,
  $n(\omega_p)=\sqrt{\epsilon(\omega_p)}$ is the linear refractive index of
  the nonlinear medium at the frequency $\omega_p$, and
  $k_p=n(\omega_p)\,\omega_p/c$ the corresponding
  wavevector~\cite{Boyd,ButcherCotter}.

The general solution of \eq{P_2Omega_1} is:
\begin{equation}
E_{\rm med}(z)=-\frac{4\pi}{c^2}\frac{
  \omega_p^2\,P_{NL}}{k_p^2-k_{NL}^2}\,e^{ik_{NL}z}+{\bar
  E}_{\rm med}\,e^{ik_p z}, \eqname{wave_med}
\end{equation}
where ${\bar E}_{\rm med}$ is a parameter to be determined.

The parameters ${\bar E}_{\rm med}$ and ${\bar E}_{\rm fs}\al{1,2}$ can be determined
by
imposing the appropriate boundary conditions at the interface between the
free-space and the nonlinear medium and at the cavity mirrors.

As reflections at the nonlinear medium interface have been taken as
negligible, the energy flux is conserved across the interface. This imposes
that:
\begin{eqnarray}
{\bar E}_{\rm med}(z_1)&=&\epsilon(\omega_p)^{-1/4}\,{\bar E}_{\rm fs}\al{1}(z_1) \eqname{bc1} \\
{\bar E}_{\rm med}(z_1+\ell)&=&\epsilon(\omega_p)^{-1/4}\,{\bar E}_{\rm fs}\al{2}(z_1+\ell).
\eqname{bc2}
\end{eqnarray}
Assuming for simplicity a vanishing reflection phase for the cavity mirrors,
the electric field is continuous at the three perfectly reflecting ones,
while at the fourth, output, mirror one has the boundary condition:
\begin{equation}
{\bar E}_{\rm fs}\al{1}(0)=\sqrt{1-T}\,{\bar E}_{\rm fs}\al{2}(L), \eqname{bc3}
\end{equation}
where the electric fields at $z=L,0$ are respectively the incident and
reflected fields, and $T$ is the mirror transmittivity.

By imposing the three conditions (\ref{eq:bc1},\ref{eq:bc2},\ref{eq:bc3})
onto the general solutions \eq{wave_fs} and \eq{wave_med} of the wave
equations in the nonlinear medium and in the two regions of free space, one
obtains explicit expressions for the three parameters ${\bar E}_{\rm med}$ and
${\bar  E}_{\rm fs}\al{1,2}$.

In particular, we are interested in the amplitude of the in-cavity field just
before the output mirror:
\begin{equation}
E_{\rm fs}\al{2}(L)={\bar E}\al{2}_{\rm fs}\,e^{i\omega_p L/c}=
\frac{4\pi\omega_p^2\,\epsilon(\omega_p)^{1/4}}{c^2}\frac{e^{ik_{NL}z_1}\,e^{-i\omega_p
    z_1/c}}{(1-T)^{1/2}-e^{-i\omega_p L_{\rm
      eff}/c}}\,
\frac{e^{i(k_{NL}-k_p)\ell}-1}{k_p^2-k_{NL}^2}
\,P_{NL}
\eqname{E_gener_1}
\end{equation}
the effective optical length being defined as
$L_{\rm eff}=n(\omega_p)\ell+(L-\ell)$.
Summing over all the pairs $(j,m)$, one finally obtains the expression
\eq{ElectrField0}.

From the general solution of the wave equation \eq{wave_med} inside the
medium, it is immediate to obtain the amplitude of the field generated in a
single-pass configuration.
One has to fix ${\bar E}_{\rm med}$ by imposing that the field vanishes at the
front interface $E_{\rm med}(z_1)=0$ and then evaluate the corresponding value of
the field at the exit interface:
\begin{equation}
E_{\rm med}(z_1+\ell)=\frac{4\pi \omega_p^2 P_{NL}}{c^2}
\frac{1-e^{i(k_{NL}-k_{p})\ell}}{k_p^2-k_{NL}^2}\,e^{ik_p\ell}\,e^{ik_{NL}z_1}
\end{equation}
from which the generated field in a single-pass geometry is immediately
obtained using the boundary condition \eq{bc2}:
\begin{equation}
E_{\rm fs}\al{2,sp}(L)=\frac{4\pi \omega_p^2\,\epsilon(\omega_p)^{1/4}}{c^2}
\,e^{ik_p\ell}\,e^{ik_{NL}z_1}\,e^{i(L-\ell-z_1)\omega_p/c}\,
\frac{1-e^{i(k_{NL}-k_{p})\ell}}{k_p^2-k_{NL}^2}\,P_{NL}. \eqname{E_gener_sp}
\end{equation}
The relation \eq{ElectrFieldT2} is then simply obtained by comparing
\eq{E_gener_1} and
\eq{E_gener_sp}. Note that \eq{E_gener_sp} could be obtained from
\eq{E_gener_1} simply by imposing that a full transmission of the output
mirror $T=1$.

\clearpage

\clearpage

\begin{thebibliography}{99}
\bibitem{HighFinesse}
This approach requires a finesse suffuciently high to avoid the contribution of
cavity modes close to unwanted resonances and may result in  very high finesse
cavities.
\bibitem{ExperimentalAssumptions}
The detailed spectrum of a mode locked laser depends critically on the
dispersion properties of its cavity, and eventually those of medium employed to
broaden its spectrum.
\end{thebibliography}
\end{document}